# Interplay of composition, structure, magnetism, and superconductivity in SmFeAs$_{1-x}$P$_x$O$_{1-y}$

N. D. Zhigadlo,[1,*] S. Katrych,[1] M. Bendele,[2,3] P. J. W. Moll,[1] M. Tortello,[4] S. Weyeneth,[2] V. Yu. Pomjakushin,[5] J. Kanter,[1] R. Puzniak,[6] Z. Bukowski,[1] H. Keller,[2] R. S. Gonnelli,[4] R. Khasanov,[3] J. Karpinski,[1] and B. Batlogg[1]

[1]*Laboratory for Solid State Physics, ETH Zurich, CH-8093 Zurich, Switzerland*
[2]*Physik-Institut der Universität Zürich, Winterthurerstrasse 190, CH-8057 Zürich, Switzerland*
[3]*Laboratory for Muon Spin Spectroscopy, Paul Scherrer Institute, CH-5232 Villigen PSI, Switzerland*
[4]*Dipartimento di Fisica, Politecnico di Torino, 10129 Torino, Italy*
[5]*Laboratory for Neutron Scattering, Paul Scherrer Institute, CH-5232 Villigen PSI, Switzerland*
[6]*Institute of Physics, Polish Academy of Sciences, Aleja Lotników 32/46, PL-02-668 Warsaw, Poland*

**Abstract**

Polycrystalline samples and single crystals of SmFeAs$_{1-x}$P$_x$O$_{1-y}$ were synthesized and grown employing different synthesis methods and annealing conditions. Depending on the phosphorus and oxygen content, the samples are either magnetic or superconducting. In the fully oxygenated compounds the main impact of phosphorus substitution is to suppress the Néel temperature $T_N$ of the spin density wave (SDW) state, and to strongly reduce the local magnetic field in the SDW state, as deduced from muon spin rotation measurements. On the other hand the superconducting state is observed in the oxygen deficient samples only after heat treatment under high pressure. Oxygen deficiency as a result of synthesis at high pressure brings the Sm-O layer closer to the superconducting As/P-Fe-As/P block and provides additional electron transfer. Interestingly, the structural modifications in response to this variation of the electron count are significantly different when phosphorus is partly substituting arsenic. Point contact spectra are well described with two superconducting gaps. Magnetic and resistance measurements on single crystals indicate an in-plane magnetic penetration depth of $\approx$ 200 nm and an anisotropy of the upper critical field slope of $\approx$ 4-5.

PACS number(s): 74.70.Xa, 74.62.Bf, 74.25.-q, 81.20.-n

## I. INTRODUCTION

A renewed interest to high temperature superconductors (HTS) was generated by the unexpected discoveries of Fe-based oxypnictide superconducting LaFePO ($T_c \cong 5$ K) [1] and LaFeAs(O,F) ($T_c \cong 26$ K) [2], followed by the subsequent development of homologous series of the quaternary iron pnictides (1111-$Ln$Fe$Pn$O, $Ln$: lanthanide, $Pn$: pnictogen) with $T_c$'s up to ~ 55 K. Several families of iron pnictides and iron chalcogenides were subsequently discovered (see Refs. [3, 4] and references therein). The compounds belonging to the 1111 family of Fe-based HTS have a common layered structure composed of an alternating stack of $Ln$O and Fe$Pn$ layers. In all series the Fe$Pn$ layer is directly responsible for superconductivity, whereas the perovskite-like $Ln$O layers play the role of a charge-carrier source. Superconductivity occurs through chemical substitution at different atomic sites of the crystal structure, or by applying an external pressure when the antiferromagnetic state is partially or completely suppressed. The resulting electronic phase diagrams depend on the particular compound [4]. One of distinctive features of 1111 family of Fe-based HTS is the possibility to induce superconductivity by applying "chemical pressure" through a partial substitution of the isovalent smaller P ions for the bigger As. Several theoretical [5] and experimental [6] studies were addressed to this issue. The consensus is that P substitution for As does not result in significant changes of the electron density, but show a clear influence on the localization of hybridized states, bandwidth, and the topology of the Fermi surface. Interestingly, having the same number of electrons and holes for any $x$ value (As$_{1-x}$P$_x$) of various parent compounds one can reliably tune the magnetic character without changing the charge carrier concentrations. However, very different responses on the P substitution were observed in a variety of systems. Upon P substitution superconductivity appears in BaFe$_2$As$_2$ [7], EuFe$_2$As$_2$ [8], CaFe$_2$As$_2$, SrFe$_2$As$_2$ [9] and LaFeAsO [6] as the tetragonal to orthorhombic structural transition temperature is lowered and the associated spin-density-wave transition is suppressed. Nevertheless, no superconductivity was observed in the CeFeAs$_{1-x}$P$_x$O system down to 2 K [10]. Among many reported compounds, superconductivity in LaFePO [11] and SmFePO [12] was shown to depend sensitively on the overall composition and the synthesis conditions. Stoichiometric LaFePO and SmFePO are metallic and nonsuperconducting [11, 13]. To date there is no complete and conclusive proof for the occurrence of superconductivity in the SmFeAs$_{1-x}$P$_x$O system. In Ref. [13] the authors concluded, based on the temperature dependence of the resistivity, that the superconducting window in SmFeAs$_{1-x}$P$_x$O is very narrow, only in the range 0.5 < $x$ < 0.65 with a maximum $T_c$ of 4.1 K at the optimal doping $x$ = 0.565.

The main objectives of this study are twofold. Firstly, we investigated the possibility to induce superconductivity in the SmFeAs$_{1-x}$P$_x$O by applying chemical pressure through the substitution of P ions for As. Secondly, we explore the relationship between structure, composition, and superconducting properties in order to elucidate the microscopic structural aspects with the macroscopic occurrence of superconductivity. We have found, through an exhaustive set of measurements, that the stoichiometric samples prepared at ambient pressure are not superconducting down to 2 K just as in LaFePO [11]. The superconductivity produced in high pressure synthesized samples is due to induced oxygen deficiency, which strongly affects the intra- and interlayer spacing dimensions and the geometry of the Fe(As,P)$_4$ and SmO$_4$ tetrahedral units.

## II. EXPERIMENTAL DETAILS

Polycrystalline samples of the SmFe(As,P)O system were synthesized at ambient pressure (AP) in evacuated quartz ampoules and under high pressure (HP). Powders of SmAs, SmP, Fe$_2$O$_3$, and Fe of high purity (≥ 99.95%) were weighed according to the stoichiometric ratio, thoroughly ground, and pressed into pellets that were then placed into alumina crucibles with cups. The crucibles were loaded to the quartz ampoules, evacuated, filled with Ar gas (~ 0.2 atm.) and sealed. The ampoules were slowly heated to 660 °C where the temperature was maintained for 3 h. Then the temperature was raised to 1060 °C and maintained for 100 h, and subsequently the ampoules were quenched in cold water. For the high pressure synthesis we used the same starting materials. The mixed multiple variants of the synthesis were used as well. In order to recognize the samples synthesized in various conditions we introduced abbreviation codes describing the sequence of synthesis. For example, the abbreviation HP+AP+HP means that the sample which was initially synthesized under high pressure (HP), was resynthesized at ambient pressure (AP), and then once more synthesized under high pressure (HP). For the growth of single crystals a NaCl/KCl flux was mixed with an equal amount of the precursor (for details see Refs. [14, 15]). All procedures related with the sample preparation were performed in a glove box due to the toxicity of arsenic. In HP synthesis the sample was enclosed in a boron nitride container and placed inside a graphite heater. In a typical run, a pressure of 3 GPa was applied at room temperature. While keeping the pressure constant, the temperature was ramped up in 1 h to the maximum value of ~ 1350 °C, maintained for 4.5 h, followed by quenching. For the crystal growth the maximum temperature was maintained for 72 h, and then decreased to room temperature in 2 h. Afterwards the pressure was released and the sample removed. The NaCl/KCl flux was dissolved in water.

Powder x-ray diffraction (XRD) studies were performed at room temperature on a STOE diffractometer (Cu $K_{\alpha 1}$ radiation, $\lambda$ = 1.54056 Å) equipped with a mini-phase-sensitive detector and a Ge monochromator. Powder patterns were refined with the program FULLPROF [16]. Single crystals were studied at room temperature on a *Bruker* x-ray single-crystal diffractometer. Data reduction and numerical absorption correction were performed using the *Bruker* AXS Inc. software package [17]. The crystal structure was determined by a direct method and refined on $F^2$, employing the SHELXS-97 and SHELXL-97 programs [18]. The magnetization measurements were carried out with a *Quantum Design* Magnetic Property Measurement System (MPMS-XL). Four-point resistivity measurements were performed in a 14 Tesla *Quantum Design* Physical Property Measurement System (PPMS). Micrometer-sized Pt leads were precisely deposited onto a plate-like crystal using a focused ion beam (FIB) method without altering the bulk superconducting properties [19]. Point-contact Andreev-reflection (PCAR) spectroscopy measurements were performed at the Politecnico di Torino and the details of the experimental set up were described in previous reports [20]. Zero-field (ZF) and longitudinal-field (LF) muon spin rotation (µSR) experiments were performed at the πM3 beam line at the Paul Scherrer Institute (Switzerland). In the ZF experiments the muons probe the internal field distribution in the samples whereas LF experiments provide information whether the internal field is static or dynamic. Neutron powder diffraction experiments were carried out at the SINQ spallation source of Paul Scherrer

Institute using the high-resolution diffractometer for thermal neutrons HRPT [21]. Since the Sm has very large absorption cross section (5920 barn) we used a double wall vanadium container with 9 mm outer and 8 mm inner diameters to reduce the absorption effect. Calculated attenuation length amounted to 0.5 mm for $\lambda = 1.494$ Å, assuming the packing density 2g/cm$^2$. The double wall container gives only 1 mm of the neutron path making this neutron experiment doable. The sample was rotated during the measurements in order to minimize aberrations due to a non-uniform packing density. The refinements of the crystal structure were carried out with the program FULLPROF [16]. The following scattering lengths were used: Sm 0.8-$i$1.65 fm, Fe 9.45 fm, P 5.13 fm, As 6.58 fm, O 5.803 fm.

## III. RESULTS AND DISCUSSION

### A. Ambient pressure vs high pressure synthesis

Already at the beginning of our exploratory synthesis study we have recognized a striking difference in the phase purity and magnetic response of the samples obtained at AP and HP conditions. Phosphorus substituted SmFeAsO samples synthesized at AP are single phase, but nonsuperconducting, whereas the HP prepared samples are less pure but show some diamagnetic response. This motivated us to undertake a systematic study. Figure 1 depicts XRD patterns of polycrystalline SmFeAs$_{1-x}$P$_x$O samples with nominal phosphorus content of $x = 0.0$, 0.5, and 1.0 synthesized at AP and under HP, and combinations of AP and HP treatments. All peaks in the XRD patterns after AP heat treatment can be indexed based on a tetragonal unit cell (*P4/nmm*) of the ZrCuSiAs-type structure. No superconductivity above 2 K was observed for those samples. In the next step, the sample with nominal composition of SmFeAs$_{0.5}$P$_{0.5}$O was heat treated under HP (~ 1350 °C, 3 GPa). This method produces highly dense samples, while the samples synthesized at ambient pressure are not dense, probably due to the low synthesis temperature (1060 °C). In the XRD pattern of the SmFeAs$_{0.5}$P$_{0.5}$O sample (abbreviated as AP+HP in Fig. 1) besides the main peaks belonging to the 1111 phase, a few additional tiny peaks were also evident that could be ascribed to traces of Sm$_2$O$_3$ and SmAs. Thus, the real composition of the polycrystalline samples after HP treatment deviates slightly from the nominal composition, as we also quantify later in the structure analysis. Surprisingly, the magnetization measurement revealed a magnetic response ascribed to superconductivity with $T_c$ = 10.2 K. Realizing the importance of high pressure treatment in inducing superconductivity additional synthesis experiments were performed.

When the SmFeAs$_{0.5}$P$_{0.5}$O sample was directly synthesized from starting component at HP, superconductivity appeared at $T_{c,\text{eff}}$ = 23.4 K (Fig. 2(a)). XRD revealed in these samples the presence of Sm$_2$O$_3$ and SmAs in a slightly higher amount compared to the AP+HP (Fig. 1). The left inset in Fig. 2(a) displays the temperature dependence of the electrical resistivity (ρ) of SmFeAs$_{0.5}$P$_{0.5}$O HP sample. The resistivity exhibits metallic character before the onset of the superconducting transition. The room temperature-to-residual resistance ratio is ≈ 3. The transition width is rather sharp, suggesting homogeneous nature of the sample.

In the next step, the HP SmFeAs$_{0.5}$P$_{0.5}$O sample was sealed in quartz ampoule and heat treated again in the same manner as it was done for AP synthesized samples. A similar, but vice versa effect is observed, *i.e.*, after HP+AP treatment the samples become practically single phase (Fig. 1, HP+AP) but nonsuperconducting. Through again applying high pressure treatment (HP+AP+HP) superconductivity is recovered but now with lower $T_c$ = 18.0 K and the sample has a less amount of impurity phases.

Based on these observations we conclude that the superconductivity appears only in the non-stoichiometric samples after heat treatment under HP. Phosphorus substitution for As itself does not influence the charge doping but non-stoichiometry, for example oxygen deficiency can introduce electrons into FeAs/P conducting layer. To check this idea further we have attempted to induce charge carriers through introducing the oxygen deficiency. The sample with nominal composition of SmFeAs$_{0.5}$P$_{0.5}$O$_{0.85}$ was synthesized in a quartz ampoule in the same way as the previous samples. To avoid/minimize possible oxygen contamination from the ampoule, the pellet was completely covered by NaCl powder. No diamagnetism was observed down to 2 K for this sample. Subsequently the sample was divided into two parts: one was heat treated at HP and the second part was sealed, together with a piece of Ta foil, in an evacuated (~ 10$^{-5}$ Torr) quartz ampoule (process abbreviated as AP$_{\text{vac}}$). The HP annealed sample is again superconducting with $T_{c,\text{eff}}$ = 16.0 K (Fig. 2(b)) and contains some impurity phases, *i.e.* $Sm_2O_3$ and SmAs. In contrast the sample $AP_{\text{vac}}$, annealed for 3 h at 1060 °C, still was nonsuperconducting. The right inset in Figure 2(b) shows the temperature-dependent resistivity for the $AP_{\text{vac}}$ and AP+HP samples. The room-temperature resistivity is larger in $AP_{\text{vac}}$ than that of the AP+HP. No trace of superconductivity is observed in the $AP_{\text{vac}}$ sample, although the SDW transition is fully suppressed, whereas a superconducting transition at $T_c \sim$ 18 K with zero resistivity is observed in AP+HP sample.

Following the results of Ref. [13] we have synthesized $SmFeAs_{0.435}P_{0.565}O$ samples at AP condition and no superconductivity was detected down to 2 K. Furthermore we have attempted to induce superconductivity in nonsuperconducting $SmFeAs_{0.435}P_{0.565}O$ sample by applying hydrostatic pressure. For this measurement a miniature container of CuBe was employed as a pressure cell and a mixture of mineral oil and kerosine was used as a pressure-transmitting medium. The pressure at low temperatures was determined by the pressure dependence of the superconducting transition temperature of pure tin placed near the sample. No trace of superconductivity for the $SmFeAs_{0.435}P_{0.565}O$ sample was found under hydrostatic pressure of 1.3 GPa in the temperature range down to 1.9 K. *None of the As free SmFePO samples show any indications of superconductivity down to 2 K whether they were AP and AP+HP prepared.*

All these results suggest that the superconductivity in the SmFe(As,P)O system occurs only after a high pressure treatment, and thus superconductivity may be related to either non-stoichiometry and/or oxygen deficiency. High-pressure preparation may induce sufficient oxygen deficiency and an increased electron concentration sufficient to realize the superconductivity, while annealing in vacuum apparently does not produce enough oxygen deficiency to obtain superconductivity. An even higher oxygen deficiency could lead to decomposition of the compound.

**B. Structure modifications due to substitution of P for As**

The changes of the crystal structure caused by As/P substitution and the structural modifications due to pressure-induced oxygen deficiency were studied by means of x-ray and neutron diffraction. In the P substituted samples the XRD peaks shift toward a higher $2\theta$ value. The (00l) peaks shift with P substitution much more than the (hk0) peaks, indicating that the *c*-axis shrinks more than the *a*-axis (Fig. 1). The lattice constants, atomic position parameters, and selected bond lengths and angles obtained from polycrystalline samples after Rietveld refinement are summarized in Table 1. The resulting As/P occupation obtained from x-ray refinement were further confirmed by EDX analysis. For easy comparison, the structural response to P substitution and to the HP treatment is shown in Fig. 3. It is evident that the changes in the bond length, angle, and layer thickness are significant (see Table 1 and Fig. 3). The Fe-As distance decreases linearly upon increasing *x*, as expected for P substitution [22], and the As-Fe-As ($\alpha$) angle increases linearly. Compared to the unsubstituted SmFeAsO sample ($a$ = 3.92520(4) Å; $c$ = 8.4693(1) Å), both lattice parameters changed to $a$ = 3.90422(4) Å and $c$ = 8.3206(1) Å for 50 % P substitution (AP method); *i.e.* $\Delta a/a \approx$ -0.53 % and $\Delta c/c \approx$ -1.76 %. The unit cell volume difference between SmFeAsO and SmFePO is about 5.8 %.

The Rietveld refinement of the structural parameters of the SmFeAs$_{1-x}$P$_x$O samples prepared at AP reveals an almost monotonic decrease in the *a* and *c* lattice parameters (and unit cell volume) with increasing P substitution. However, the response of the lattice metrics is strongly anisotropic with the interlayer spacing showing a significantly larger contraction than the change in basal plane dimensions. These values clearly reveal the diversity in bonding, with less contraction in the covalently bonded Fe-As/P layers. The interlayer contraction is ~ 2.5 times larger, consistent with weaker interlayer interactions. The reduction along the *c*-axis is due to the contraction of the As/P-Fe-As/P layer thickness (*S*2 in Fig. 3(a)), *i.e.* a reduction of the FeAs/P height (Fig. 3(c)) when arsenic is replaced by the smaller phosphorus, while the Sm-As(P) distance and SmO layer thickness remain essentially unchanged with increasing P substitution (Table I).

These trends are different from that in SmFeAsO$_{1-x}$F$_x$, Sm$_{1-x}$Th$_x$FeAsO, and SmFeAsO$_{1-x}$ [23], where the *c*-axis lattice constant contraction is due to a large reduction of the Sm-As distance, while the Sm-O and As-Fe-As layers thickness actually increase with increasing doping. These fundamental differences reflect the fact that F, Th, and O deficiency brings the Sm-O layer closer to the As-Fe-As block and facilitates electron transfer, while P substitution in AP samples is a pure geometrical lattice effect without charge carrier transfer. We note that the Fe-As distance (~ 2.40 Å in Sm-1111, for example) is essentially doping independent (see SmFeAsO$_{1-x}$F$_x$, SmFeAsO$_{1-x}$, Sm$_{1-x}$Th$_x$FeAsO [23]) but decreases rapidly with increasing P substitution. The reduced Fe-As/P distance may modify the hybridization between the Fe 3*p* and the As 4*p* orbitals and thus quench the ordered Fe magnetic moment.

**C. Structure modifications due to doping by oxygen deficiency**

Next we discuss some evidence that a variable oxygen content reflects itself in subtle modifications of the structure. As noticed in previous studies [24] oxygen deficiency in the *Ln*-O layer has minute influence on the *Ln*-O layer geometry (bond length), but the associated increase in the charge transferred to the Fe-*Pn* layer causes substantial modification of the Fe-*Pn* bonding geometry. Early theoretical calculations [25] and recent experimental studies [26] reveal exceptionally strong dependency of the P-P bond length on the electron count. Remarkably, the Fe-P reacts differently from the Fe-As, because of occupation of anti-bonding P-P states [26]. Qualitatively, the structure variations in this study appear to follow the same trends. We describe that the anomalous changes in the shape of the Fe(As,P)$_4$ tetrahedra is driven by an unusually high sensitivity of the bond length involving P to the total electron count.

In our SmFeAs$_{0.5}$P$_{0.5}$O HP samples superconductivity sets in when the Sm-O bond length becomes longer (~ +0.004 − 0.005 Å) and Sm-As/P shorter (~ -0.016 − 0.023 Å), compared with that in nonsuperconducting SmFeAs$_{0.5}$P$_{0.5}$O prepared at AP (Table I). The oxygen deficiency in the samples can be estimated as follows. From the neutron diffraction data of LaFeAsO$_{1-y}$ and NdFeAsO$_{1-y}$ a universal relationship between bond length and oxygen deficiency was established (Fig. 5 in Ref. [24]): *an increase by ~ 0.001 Å in Ln-O and a decrease by ~ 0.04 Å in Ln-As bond lengths corresponds to ~ 1 % oxygen deficiency*. Accordingly our HP samples contain ~ 4 − 6 % O-deficiency. Oxygen deficiency make more electrons available for transfer to the Fe-As/P layer, modifying the bonding geometry in Fe(As,P)$_4$ and shifting the Fermi level to higher energies. Again, in line with the difference between the Fe-As and Fe-P response to the changing the Fermi level [26], we notice a similar difference between SmFeAs(O,F) and the present compounds with P. As mentioned above, FeAs$_4$ and Fe(As,P)$_4$ respond differently to the changes in the electron count. In Ref. [26] the electron count was changed by substitution of P for Ge, while here we modify it through the chemical composition of Sm-O layer. In SmFeAs(O,F) [14, 15] and (Sm,Th)FeAsO [23] the FeAs$_4$ geometry remains almost unchanged upon F and Th substitution (shaden area in Fig. 4), while in the present study we found a significant change of the Fe(As,P)$_4$ geometry upon O deficiency in the Sm-O layer. Figure 3(b) and 3(c) shows the influence of pressure-induced oxygen deficiency on the interlayer distance ($S_3$) and pnictogene height ($h_{Pn}$). These structural trends were checked through various heat treatments and all HP synthesized superconducting samples clearly show different structural response, *i.e., the oxygen deficiency weakens the bonding in the SmO layer, brings it closer to conducting FeAs/P layer and subsequently changes the Fe(As/P)$_4$ tetrahedron in the direction of $h_{As,P}$ elongation* (Fig. 3).

A direct search of the oxygen deficiency is a difficult task. We have applied high resolution powder neutron diffraction technique to address this issue. AP nonsuperconducting and HP superconducting samples with the same nominal composition of SmFeAs$_{0.5}$P$_{0.5}$O were synthesized for neutron diffraction studies. Figure 5 shows a neutron diffraction pattern and its refinement for the AP SmFeAs$_{0.5}$P$_{0.5}$O sample. The increase in intensity at large 2$\theta$ is typical for strongly absorbing materials. Fortunately, the structure does not have many parameters, $z$-Sm, $z$-P, and oxygen occupancy. All atomic displacement parameters (ADP) were fixed by the literature data for LaFePO taken from Ref. [11]. Both samples were refined with the same varying parameters, namely $z$-P, oxygen occupancy, lattice constants, $U$-half width parameter of the Cagliotti-type of the resolution function, 11 background Fourier coefficients, overall ADP and sin2$\theta$ systematic line shift parameter to account for the absorption. Only the $z$-Sm was fixed because Sm has a very small scattering length and does not substantially contribute to scattering. The lattice constant $c$ = 8.33705(63) Å for the HP sample is substantially smaller then that of the AP sample ($c$ = 8.35325(27) Å), whereas $a$ stays the same within the error bars (HP, $a$ = 3.91275(30) Å; AP, $a$ = 3.91244(11) Å). However, $z_{As/P}$ is larger in the HP sample than that in AP sample ($z_{As/P}$ = 0.65397(127) for HP and $z_{As/P}$ = 0.64854(65) for AP), resulting in a larger Fe-As/P inter-plane spacing ($h_{As/P}$), that is in line with the higher $T_c$. The line broadening is almost two times higher in the HP sample implying that the sample contains more micro strain (dispersion of lattice constant, or inhomogeneity). The sample synthesized under HP clearly shows oxygen deficiency since its occupancy is smaller than unity (0.85(4)), whereas for AP sample the refined value is 1.07(3). With this we conclude that superconductivity in SmFeAs$_{1-x}$P$_x$O can be related with the oxygen deficiency.

**D. Muon spin rotation studies: static and dynamic fields**

The polycrystalline SmFe(As,P)O samples were investigated by muon-spin rotation (µSR) experiments in zero field (ZF), weak-transverse field (WTF), and longitudinal field (LF). Here LF and TF denote the cases where the applied magnetic field is parallel and perpendicular to the initial muon-spin polarization, respectively. In WTF experiments the muons stopping in magnetically ordered regions of the samples lose their polarization relatively fast, because the field at the muon stopping site is a superposition of the high internal field and the weak applied external field. For a detailed description of the µSR technique see e.g. Ref. [27].

The static or quasi-static magnetic response of the samples was analyzed by using the following functional form [27]:

$$P^{ZF}(t) = P_{\text{stat}} \left( \frac{2}{3} f_{\text{osc}} \exp[-\lambda_t t] + \frac{1}{3} \exp[-\lambda_l t] \right). \quad (1)$$

Here the 2/3 oscillating (1/3 nonoscillating) component is caused by the internal fields at the muon stopping site which are transverse (longitudinal) to the initial muon-spin polarization. $\lambda_t$ and $\lambda_l$ are the corresponding exponential depolarization rates. The oscillating component takes the form $f_{\text{osc}} = \cos(\omega_0 t + \phi_0)$ or $f_{\text{osc}} = J_0(\omega_0 t + \phi_0)$, depending on the samples. Here, $\omega_0$ is the precession frequency, $J_0$ is a zeroth-order Bessel function, and $\phi_0$ the phase, that is fitted close to zero in all cases. The cosine and the Bessel functions point towards the commensurate and the incommensurate magnetic order, respectively [28].

As a first step we describe the SmFe(As,P)O samples prepared by the HP method. In figure 6(a) the oscillating muon-time signal of unsubstituted SmFeAsO reveals that due to the fact that the oscillating part of the muon-time signal is described by a simple cosine function, the SmFeAsO exhibits static commensurate magnetic order. The Néel temperature $T_N$ = 120 K (see Fig. 7) is reduced compared to earlier measurements of SmFeAsO synthesized at ambient pressure ($T_N$ = 135 K [29]). Such a difference likely occurs due to slightly reduced oxygen content. As shown above this effect could result from the high pressure synthesis. Upon substituting 30% of As by P the sample still exhibits static magnetic order, yet it is described by a Bessel function (Fig. 6(b)). Both the ordering temperature $T_N$ and the frequency are suppressed compared to the unsubstituted compound (see Fig. 7).

The superconducting SmFeAs$_{0.5}$P$_{0.5}$O sample prepared under high pressure shows no oscillations in the ZF muon-time spectra (see Fig. 6(d)). Instead a fast decay of the muon-spin polarization is observed at low temperature which could indicate that the magnetism becomes dynamic in the muon time scale in this sample. In fact, evidence for dynamic magnetism follows from LF measurements. The data for both ZF and LF measurements are best described by a superposition of two exponentially decaying functions. This is in agreement with earlier studies in 1111 compounds and theoretical calculations of the muon stopping site, which revealed two distinct stopping sites of the muons in the 1111 compounds, one close to the Fe atoms in the FeAs/P layers and one near the LaO planes [30, 31]:

$$P^{\text{ZF, LF}}(t) = P_{\text{fast}} \exp[-\lambda_{\text{fast}} t] + P_{\text{slow}} \exp[-\lambda_{\text{slow}} t] \quad (2)$$

Here $P_{\text{fast}}$ ($P_{\text{slow}}$) and $\lambda_{\text{fast}}$ ($\lambda_{\text{slow}}$) are the part of the muon-spin polarization and the depolarization rate of the fast (slow) relaxing component, respectively. The temperature dependences of the relaxation rates $\lambda_{\text{fast}}$ and $\lambda_{\text{slow}}$ for both the ZF and LF measurements are shown in Fig. 8(a). Interestingly, up to $\mu_0 H$ = 0.64 T both the ZF and the LF relaxation rates coincide within almost the whole temperature region. Above $T$ = 80 K the relaxation rates are only weakly temperature dependent up to 300 K and the data could be also described by a single exponential decay. Below $T$ = 80 K both $\lambda_{\text{fast}}$ and $\lambda_{\text{slow}}$ increase with decreasing temperature whereas they tempt to saturate below $T$ = 40 K to a value of $\lambda_{\text{fast}}$ = 2 µs$^{-1}$ and $\lambda_{\text{slow}}$ = 0.2 µs$^{-1}$. Only below $T$ = 10 K a further increase in both relaxations is seen again.

In SmFePO the ZF (and LF) muon time spectra appear similar to those in the superconducting SmFeAs$_{0.5}$P$_{0.5}$O sample (see Fig. 6(e)). Again the data are best described with Eq. (2). Also the temperature dependence of $\lambda_{fast}$ and $\lambda_{slow}$ is very similar to that in SmFeAs$_{0.5}$P$_{0.5}$O whereas only the values of $\lambda_{fast}$ and $\lambda_{slow}$ are smaller. The fact that both the ZF and the LF relaxation rates coincide at all temperatures and the exponential character of the muon polarization decay points to the existence of fast electronic fluctuations within the µSR time window for SmFePO and SmFeAs$_{0.5}$P$_{0.5}$O prepared by high pressure synthesis (Fig. 8(a) and 8(b)). A similar temperature dependence for $\lambda_{fast}$ and $\lambda_{slow}$ was already observed in superconducting SmFeAsO$_{0.85}$ [32]. It was suggested that the fluctuations most probably are not related to superconductivity [32] which the presented data confirm, as the fluctuations are present in both the superconducting SmFeAs$_{0.5}$P$_{0.5}$O and the nonsuperconducting SmFePO. Furthermore the *T*-dependence for both samples is very similar and no feature is seen close to $T_c$. The increase of the relaxations below $T = 10$ K is most probably associated with additional local field broadening due to the ordering of the Sm moments at low temperatures [32-34].

However, the muon experiments reveal that the samples prepared initially at low pressure (AP) and those prepared at high pressure and then annealed at ambient pressure (HP+AP) change their magnetic behavior significantly. This is clearly seen by comparing the SmFeAs$_{0.5}$P$_{0.5}$O prepared at high and low pressure. While the HP is superconducting, the AP sample is nonsuperconducting and even shows static magnetic order below $T_N = 60$ K (see Fig. 7). Comparison of the ZF muon time spectra shows the difference immediately (Figs. 6(c) and 6(d)). The HP sample shows a fast decay of the muon spin polarization pointing to fast fluctuations (Fig. 6(d)) whereas the AP sample shows oscillations indicating static magnetic order (Fig. 6(c)). Thus the muon time spectra of the ZF measurements were described again by Eq. (1) whereas the magnetic signal was described by a Bessel function.

The comparison of the two SmFeAs$_{0.5}$P$_{0.5}$O samples shows that the preparation procedure (oxygen stoichiometry, geometry of the Fe(As,P) layer) influences the electronic properties in an essential way: one method produces superconducting samples with fluctuating magnetism, the other gives nonsuperconducting samples with static magnetic order.

**E. Point-Contact Andreev-Reflection spectroscopy measurements**

Point-contact Andreev-reflection (PCAR) spectroscopy measurements have been performed on the HP SmFeAs$_{0.5}$P$_{0.5}$O sample. PCAR spectroscopy, like Scanning Tunnelling Spectroscopy (STS), allows determining the number, value, and symmetry of the superconducting energy gaps in superconductors, a fundamental information necessary to unravel the microscopic mechanism at the origin of superconductivity. However, while STS can be performed only on very flat surfaces of single crystals, PCAR spectroscopy can be reliably used to investigate polycrystalline samples as well.

Instead of using the standard, classical technique where a sharp metallic tip is pressed against the sample under study, we used the so-called "soft" technique, *i.e.* the contact is created by means of a small drop of Ag conducting paste [20]. In this way, which has proven very useful in studying several materials including Fe-based superconductors [35-37] no pressure is applied to the sample and much more thermally stable contacts are achieved. The experimental conductance curves have been obtained by numerical differentiation of the *I-V* curves and have then been normalized, *i.e.* divided by their normal-state curve (at $T = T_c$), and fitted to a modified [38] two-band, *s*-wave Blonder-Tinkham-Klapwijk (BTK) model [39] generalized to take into account the angular distribution of the current injection at the interface [40]. In the two-band case, the model is simply the weighed sum of two BTK contributions [37]:

$$G(E) = w_1 G_1^{BTK}(E) + (1-w_1)G_2^{BTK}(E), \quad (3)$$

where $w_1$ is the weight of band 1. Variable $G_i$ is defined by three parameters: the gap value, $\Delta_i$, the broadening parameter, $\Gamma_i$ and $Z_i$, which accounts for the height of the potential barrier at the interface and for the mismatch of the Fermi velocities between the normal metal and the superconductor.

Figure 9 shows the temperature dependence of the normalized conductance curves (symbols) for the Ag/SmFeAs$_{0.5}$P$_{0.5}$O point contact with their relevant two-band BTK fittings curves (solid lines). The one-gap model is far from reproducing the experimental curves while a two-gap one fits remarkably well the experiment. From the fit we get 3.5 meV for the small gap and 11.5-11.9 meV for the larger one with $2\Delta/k_B T_c$ ratio of about 4 (close to the BCS value) and 13, respectively. Although the ratio for the large gap is very high and has still to be fully understood, the two-gap model fits quite well the data while the single-gap one disregards a large portion of the experimental curve and would give a $2\Delta/k_B T_c$ anomalously high (from about 6 to 8) for a one-gap picture.

The Andreev-reflection features gradually decrease in amplitude with increasing temperature until they completely disappear at the critical temperature of the contact ($T_c$ = 20 K, very close to the bulk one) and the two-band model can follow reasonably well the evolution of the curves with increasing temperature. The behavior of the two gaps, as obtained from the fit, is shown in the inset to Fig. 9 (symbols): the small gap follows rather well a BCS trend (red solid line) while the large one apparently decreases almost linearly above $\approx T_c/2$. In the same range, however, the uncertainty on the large gap is rather big so that a BCS trend (blue solid line) for this gap cannot be ruled out.

The above results seem to indicate the presence of two nodeless gaps in the SmFeAs$_{0.5}$P$_{0.5}$O compound, in agreement with what was obtained by PCAR spectroscopy measurements on other Fe-based superconductors of the 1111 [36, 37] but also of the 122 family [35, 41]. In particular, nodeless superconductivity in this compound is expected also according to the theoretical work by Kuroki *et al.* [42]: given indeed the lattice constants ($a$ = 3.90664(2) Å; $c$ = 8.3489(4) Å) and the pnictogen height ($h_{Pn}$ = 1.319(8) Å), by looking at the upper panel of Fig. 19 in Ref. [42] it is possible to notice that our SmFeAs$_{0.5}$P$_{0.5}$O lies in the "high $T_c$, nodeless" region. In this regard, our results may support the picture given in Ref. [42], *i.e.* that lattice structure considerably affects the Fermi surface and, as a consequence, the gap functions in these compounds.

## F. Magnetic and transport properties of single crystals

For the measurements presented here we chose single crystals from the batches with SmFeAs$_{0.5}$P$_{0.5}$O and SmFeAs$_{0.5}$P$_{0.5}$O$_{0.85}$ nominal compositions. According to XRD they are free of impurities, twins, or intergrowing crystals and show well-resolved reflection patterns, indicating high quality perfection. As a representative example, the structural refinement parameters of SmFeAs$_{0.5}$P$_{0.5}$O$_{0.85}$ single crystal with $T_c$ = 17 K is summarized in Table II.

Figure 10 shows the temperature dependence of the magnetic moment of a SmFeAs$_{0.5}$P$_{0.5}$O$_{0.85}$ single crystal, measured while increasing the temperature in a magnetic field of 1 mT after zero-field-cooling (zfc) and field-cooling (fc). The observed signal at lowest temperature is indicative for bulk superconductivity, and the sharp transition for a small doping spread in the sample. The onset is estimated to be $T_c$ ~ 17 K.

Initial magnetization curves were recorded for the SmFeAs$_{0.5}$P$_{0.5}$O$_{0.85}$ single crystal in the temperature range between 2 and 17 K. These allow us to gain information on the first penetration field $H_p$, which denotes the magnetic field above which vortices enter the sample. Figure 11(a) presents the measured magnetic moment in low magnetic fields, which show the linear decrease of $m(H)$ in the Meissner state and the following upturn in magnetization due to the entrance of vortices into the bulk. The field $H_p$ was estimated according to the procedure discussed in [43]. The quantity $(BV)^{1/2}$ is calculated from the measured magnetic moment and plotted as a function of magnetic field (see Fig. 11(b)). The sudden increase from zero is happening due to the penetration of vortices at $H_p$, where $B$ is non-zero anymore in the sample and vortices are overcoming the surface. In Fig. 11(c) we show the temperature evolution for $H_p$ for the SmFeAs$_{0.5}$P$_{0.5}$O$_{0.85}$ single crystal as extracted from Fig. 11(b). The field $H_p$ is related with the intrinsic $H_{c1}$, but is systematically lowered as compared to the latter due to the reduction of the Meissner state by the shape factor of the plate-like single crystal which unfortunately is expected to contribute significantly. Approaching the base temperature, $\mu_0 H_p(0) \approx 5$ mT, which yields an estimate for the in-plane magnetic penetration depth $\lambda_{ab}(0) \approx 200$ nm.

The crystals with various $T_c$ were used for magneto transport studies. Figure 12(a) shows the temperature dependence of the magnetoresistance from 2 K to 300 K for two SmFeAs$_{0.5}$P$_{0.5}$O crystals. It is well known for the parent SmFeAsO compound there is a drop in resistivity around 135 K due to a structural transition associated with a spin-density wave (SDW) type antiferromagnetic order. As we can see in Fig. 12(a) upon P substitution the anomaly in the resistance becomes less obvious and shifts to lower temperatures. For one crystal ρ(T) shows a strong upturn at low temperature, which could be result of charge carrier localization due to the disorder in the conducting layers as phosphorus is substituted on the arsenium site. In both samples an onset critical temperature $T_{c,on}$ of ~ 5.8 K and ~ 8 K is observed. Figures 12(b) and 12(c) show the temperature dependences of the magnetoresistance in various magnetic fields applied parallel to the ($Fe_2As/P_2$) layers ($H \parallel ab$) and perpendicular to them ($H \parallel c$). Furthermore the magnetoresistance show that magnetic moments of $Sm^{3+}$ ions order antiferromagnetically below magnetic transition of ~ 4.8 K which is practically field independent (dashed line in Figs. 12(c) and 12(d)) and as recently observed [44] the complete suppression of Sm antiferromagnetism would require fields which excess of 60 T. While structure analysis attests only one over-all structure, we nevertheless attribute the two electronic states to microscopic electronic inhomogeneity.

As one representative example of magnetoresistance data obtained on the crystals from the batches $SmFeAs_{0.5}P_{0.5}O_{0.85}$, in Fig. 13 we present the data for single crystal with $T_{c,zero}$ = 21.7 K. Magnetoresistance measurements $\rho(T, H)$ near $T_c$ for magnetic fields parallel ($H \parallel ab$) and perpendicular ($H \parallel c$) to the FeAs/P-planes show remarkably different behavior than that of Sm1111 substituted with F for O or Th for Sm [15, 23]. In later two cases the magnetic fields cause only a slight shift of the onset of superconductivity, but a significant broadening of the transition, indicating weak pinning and accordingly large flux flow dissipation. In P substituted Sm1111 crystals the presence of magnetic fields shift the onset of superconductivity to lower temperatures, but do not cause a significant broadening of the transition. The upper critical fields $H_{c2} \parallel ab$ and $H_{c2} \parallel c$ extracted from the resistivity measurements is shown in Fig. 13(b). The upper critical fields $H_{c2}$ in $SmFeAs_{0.5}P_{0.5}O_{0.85}$ increase linearly with decreasing temperature ~ 1 K below $T_c$, with a slope of ~ 5.7 T/K ($H \parallel ab$) and ~ 1.3 T/K ($H \parallel c$). The slopes in Sm1111 crystals substituted with F or Th depend strongly on the criterion due to pronounced broadening. We found 8.0 – 5.5 T/K for $H \parallel ab$ and 3.3 – 1.2 T/K for $H \parallel c$ [15, 23]. All these large slopes indicate very high values of $H_{c2}(0)$. The Sm1111 structure is more anisotropic than the structure of 122, which is already manifested in the upper critical field anisotropy $\gamma_H = H_{c2} \parallel ab / H_{c2} \parallel c$ [15]. The data presented in Fig. 13(b) suggest that the anisotropy $\gamma_H$ in $SmFeAs_{0.5}P_{0.5}O_{0.85}$ sufficiently below $T_c$ decreases from ~ 7.3 to ~ 6.6 with decreasing temperature. Such temperature dependent $\gamma_H$ further supports a multi-band superconductivity scenario, where different parts of the Fermi surface sheet develop distinct gaps in the superconducting state.

## IV. CONCLUSIONS

Polycrystalline and single-crystalline samples of SmFeAs$_{1-x}$P$_x$O were successfully prepared using quartz ampoule and the high-pressure cubic anvil techniques. In the case of ambient pressure prepared samples substitution of As by P results in a decrease in the unit-cell volume and a continuous suppression of both the ordering temperature and the frequency of magnetic order. The appearance of superconductivity in the SmFeAs$_{1-x}$P$_x$O$_{1-y}$ samples caused by the oxygen deficiency was realized only after heat treatment under high pressure. The pressure-induced oxygen deficiency brings the Sm-O charge-transfer layer closer to the superconducting As/P-Fe-As/P block and facilitates electron transfer. For superconducting SmFeAs$_{0.5}$P$_{0.5}$O$_{1-y}$ samples only dynamic magnetism survives, while the ambient prepared samples with the same amount of P substitution still show a static magnetic moment at temperatures below ~ 60 K. Point-contact Andreev-reflection spectroscopy indicates the existence of two energy gaps in superconducting samples supporting a common multigap scenario proposed for FeAs-based superconductors. Single crystals of SmFeAs$_{1-x}$P$_x$O$_{1-y}$ have been grown under high pressure and their crystallographic and basic superconducting state properties were presented. The upper critical field deduced from resistance measurements is anisotropic with slopes of ~ 5.7 T/K ($H \parallel ab$ plane) and ~ 1.3 T/K ($H \parallel c$-axis) sufficiently far below $T_\text{c}$.

**ACKNOWLEDGMENTS**

We would like to thank P. Wägli for the EDX analysis. This work was supported by the Swiss National Science Foundation and the National Center of Competence in Research MaNEP (Materials with Novel Electronic Properties), Project № 124616.

# References

*zhigadlo@phys.ethz.ch

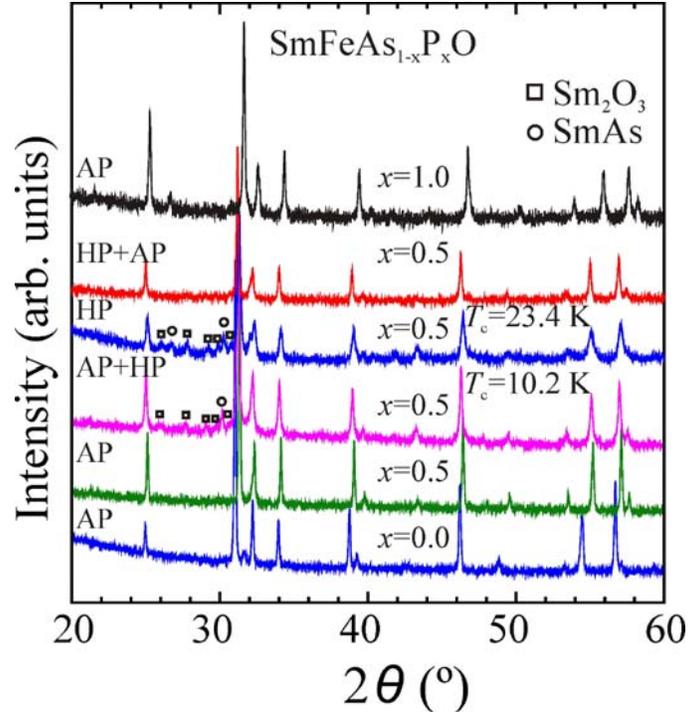

**FIG. 1.** (Color online) X-ray diffraction patterns of polycrystalline samples with nominal composition SmFeAs$_{1-x}$P$_x$O ($x$ = 0.0, 0.5, and 1.0) synthesized at ambient pressure (AP) and high pressure (HP) conditions. The AP sample with $x$ = 0.5 was heat treated at HP and abbreviated as AP+HP whereas the HP sample was heat treated at AP and abbreviated as HP+AP. The peaks marked by squares and circles belong to Sm$_2$O$_3$ and SmAs, respectively. The values of $T_c$ were determined from the magnetic susceptibility measurements.

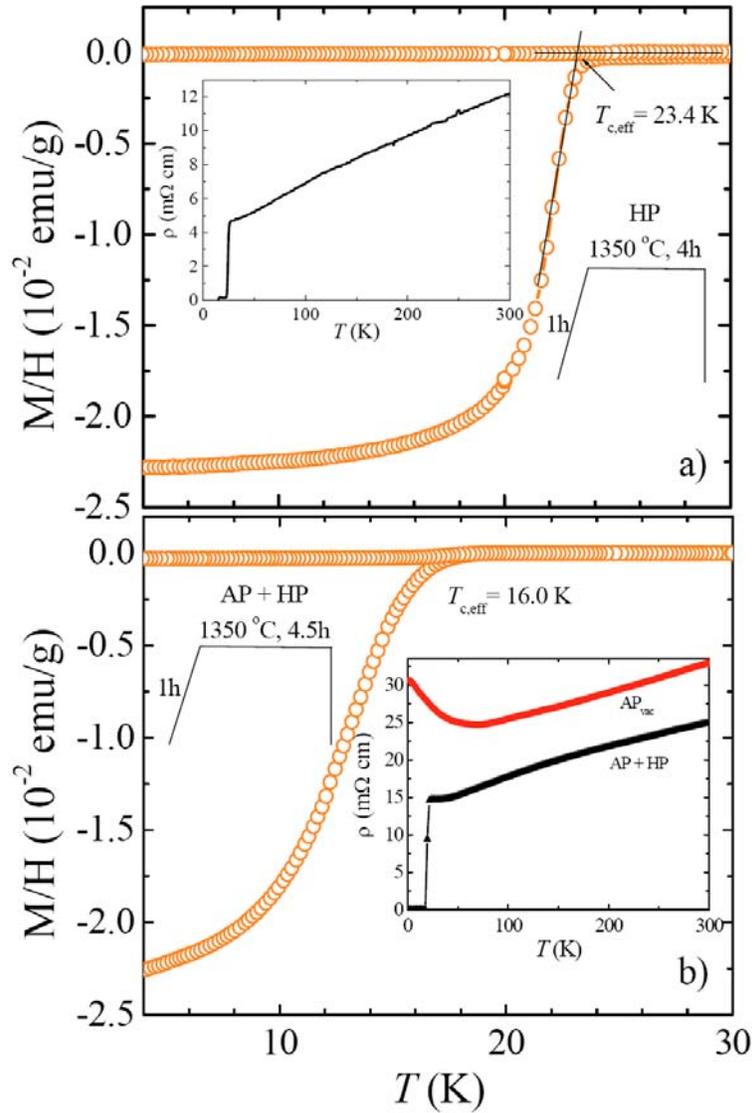

**FIG. 2.** (Color online) Temperature dependence (zero-field-cooled and field-cooled) of the magnetic susceptibility of the high pressure treated superconducting SmFeAs$_{1-x}$P$_x$O samples. The determination of $T_{c,eff}$ is illustrated. In both panels the temperature history during the high pressure synthesis processes are sketched. In the insets to Fig. 2(a) and (b) the temperature dependent resistivity for the studied samples is presented.

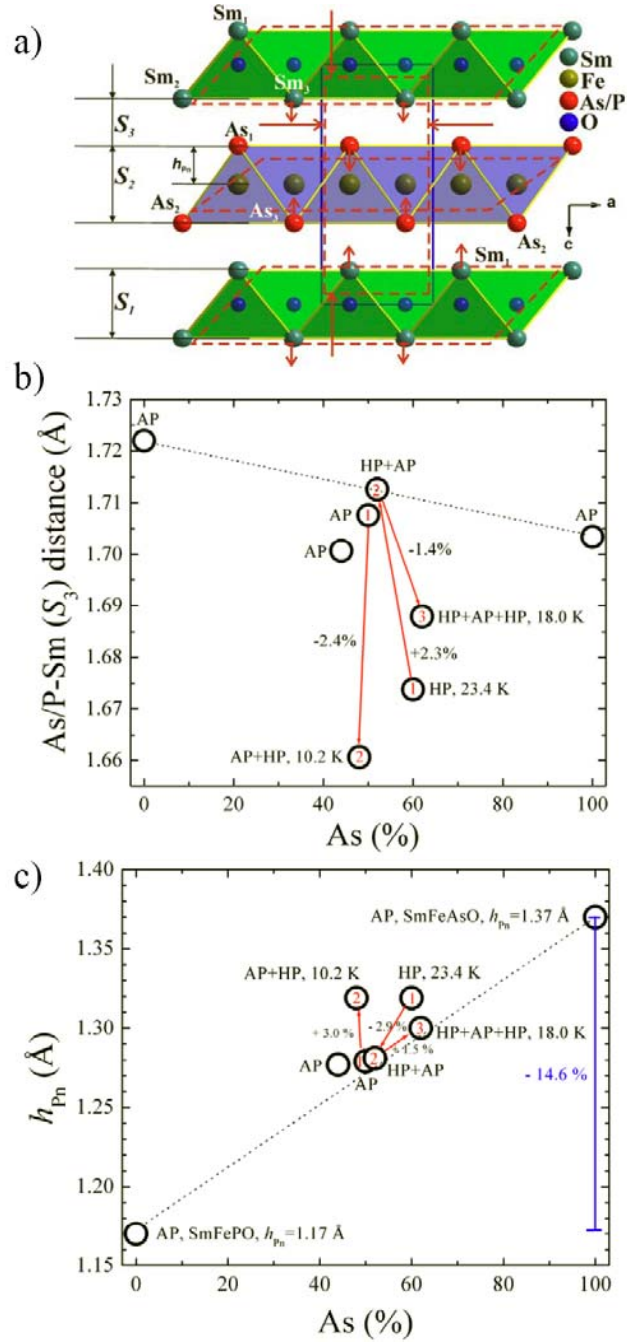

**FIG. 3.** (Color online) (a) Schematic representation of the projection of the SmFeAs$_{1-x}$P$_x$O lattice on the *ac* plane and its changes with substitution of As by P (red dotted lines). (b) Interlayer distance $S_3$ and (c) pnictogen height $h_{Pn}$ for SmFeAs$_{1-x}$P$_x$O samples synthesized at ambient pressure (AP), high pressure (HP), and their combinations. Digits in the circles show the sequence of heat treatment. All error bars are smaller than the size of symbols.

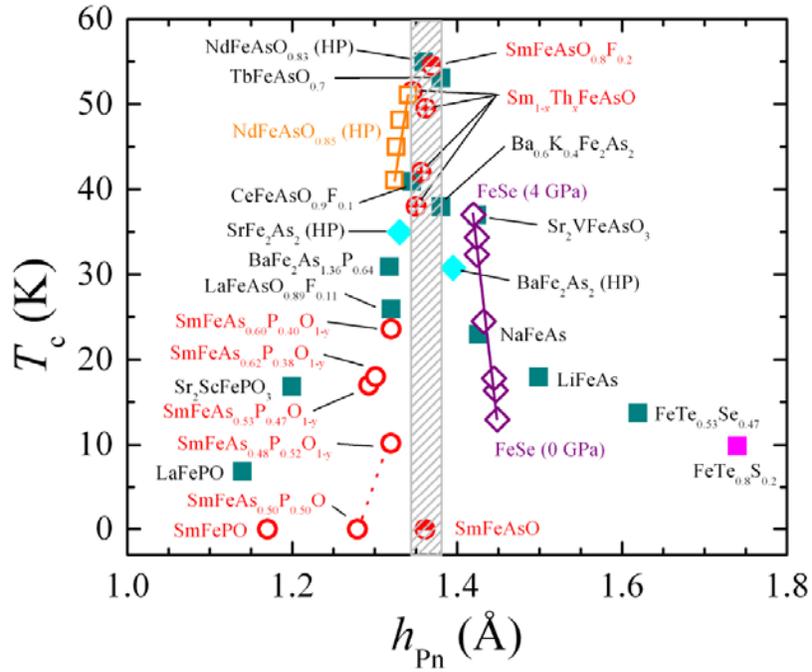

**FIG. 4.** (Color online) Anion height dependence of the superconducting transition temperature $T_c$ for various Fe-based superconductors, adopted from Ref. [45]. Filled squares indicate the data obtained at ambient pressure. Filled diamonds are the data of SrFe$_2$As$_2$ and BaFe$_2$As$_2$ obtained under optimal pressure. Open squares indicate the data of NdFeAsO$_{0.85}$ under high pressure. The data of FeSe under high pressure are indicated by open diamonds. Present experimental results and data from [14, 15, 23] are herein are indicated by red circles. The dashed line highlights the changes in $h_{Pn}$ due to oxygen deficiency.

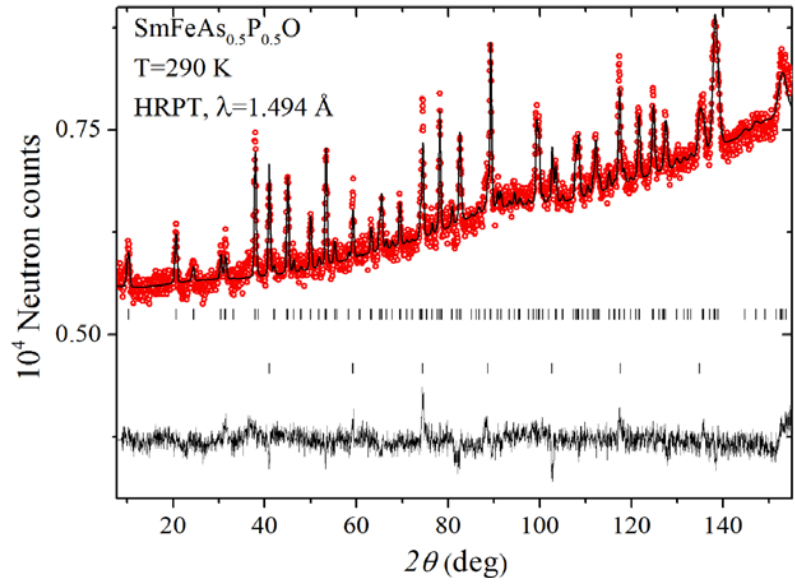

**FIG. 5.** The Rietveld refinement pattern and difference plot of the neutron diffraction data for the SmFeAs$_{0.5}$P$_{0.5}$O (AP) sample collected at 290 K using the instrument with the wavelength $\lambda$ = 1.494 Å. The rows of tics show the Bragg peak positions for the main phase and the vanadium sample container.

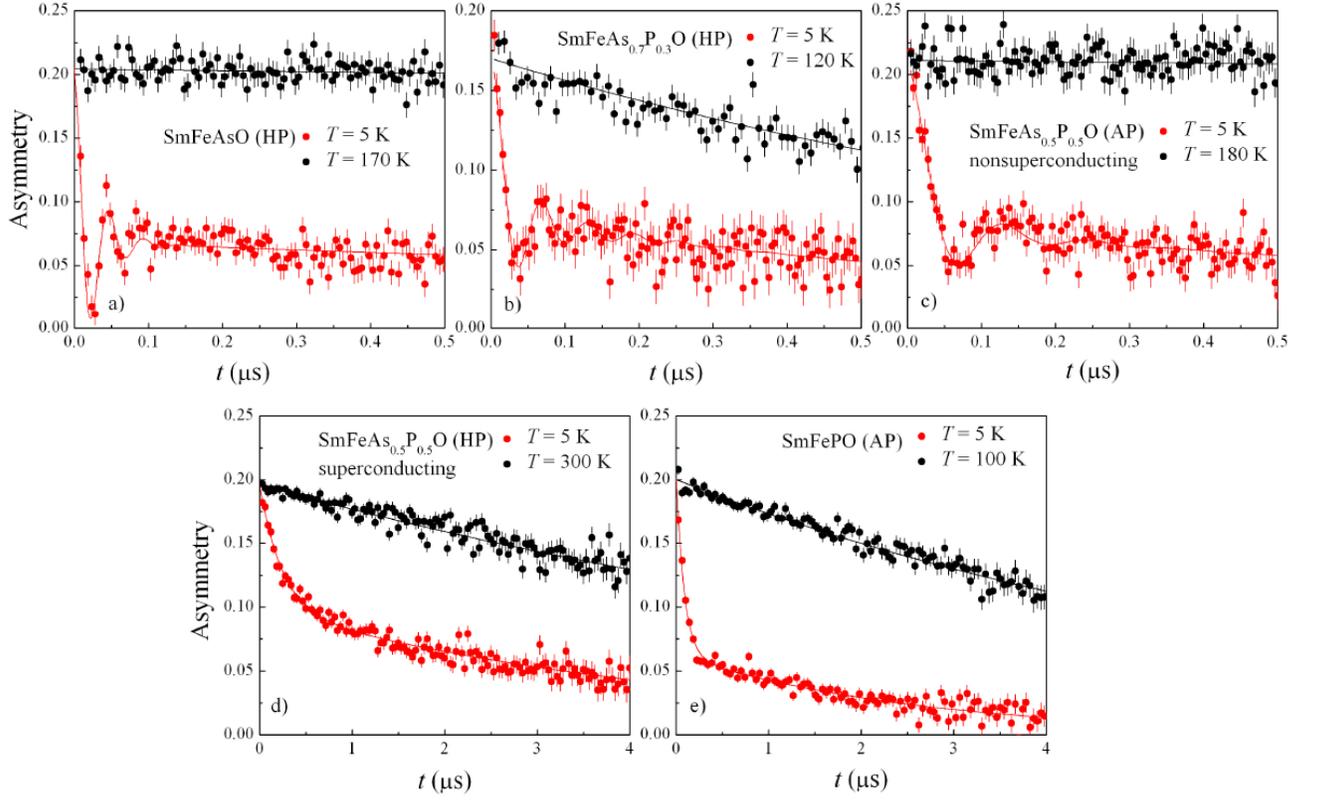

**FIG. 6.** (Color online) ZF μSR time spectra of the muon spin polarization for various dopings of SmFeAs$_{1-x}$P$_x$O at representative temperatures below and above the magnetic ordering temperature for a) SmFeAsO (HP), b) SmFeAs$_{0.7}$P$_{0.3}$O (HP), c) SmFeAs$_{0.5}$P$_{0.5}$O (AP), d) SmFeAs$_{0.5}$P$_{0.5}$O (HP), and e) SmFePO (AP).

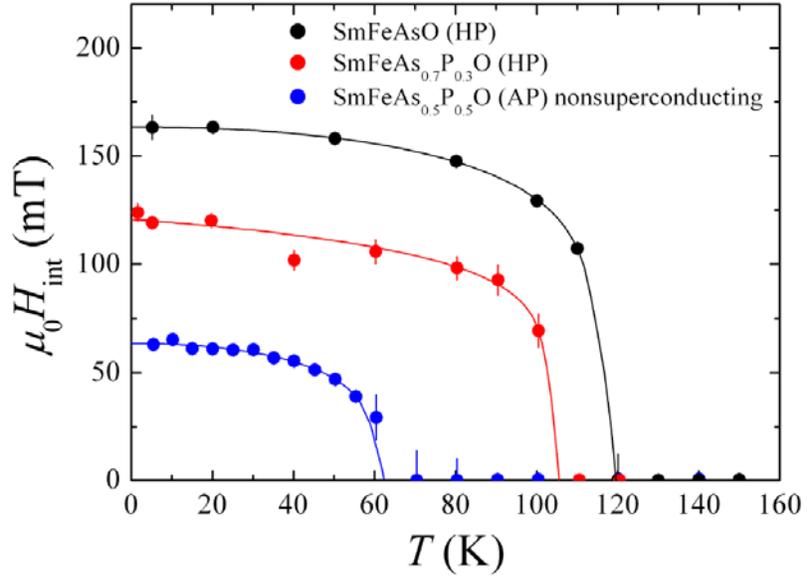

**FIG. 7.** (Color online) The local field $\mu_0 H_{int}$ as a function of substitution and temperature, deduced from the ZF μSR measurements on SmFeAs$_{1-x}$P$_x$O. The lines are guides to the eye.

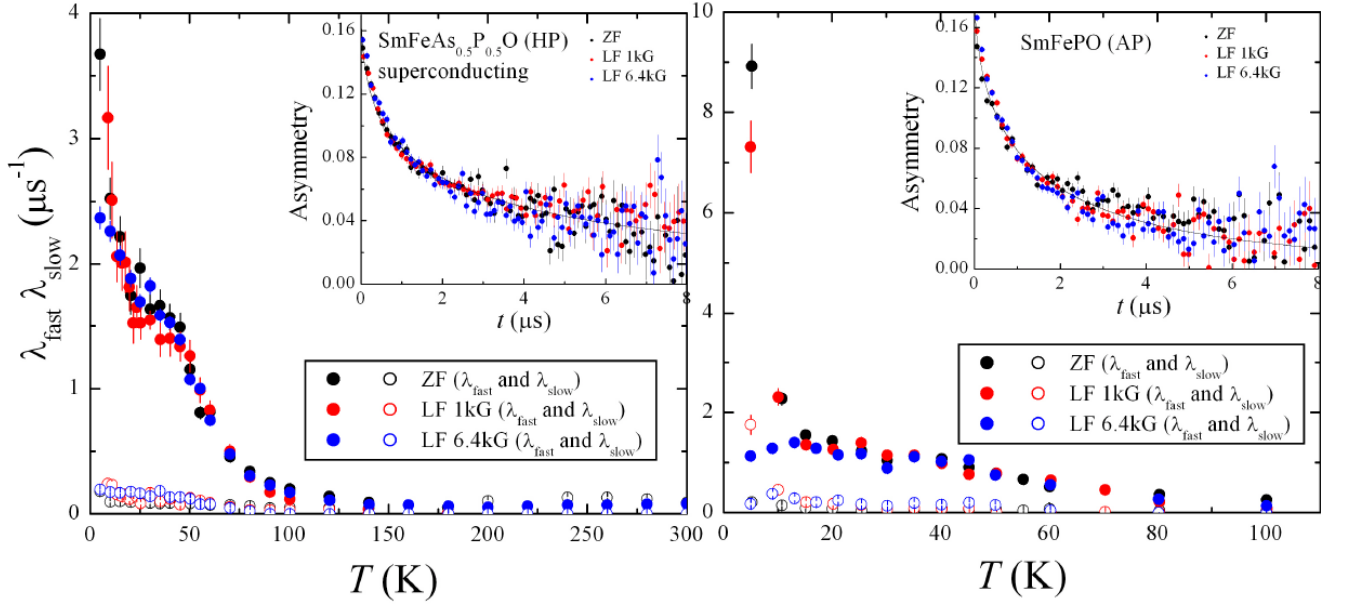

**FIG. 8.** (Color online) Temperature dependence of the muon depolarization rates $\lambda_{fast}$ and $\lambda_{slow}$ of the ZF and LF muon depolarization rates for a) SmFeAs$_{0.5}$P$_{0.5}$O (HP) and b) SmFePO (AP). The solid lines represent fits of Eq. (2) to the data.

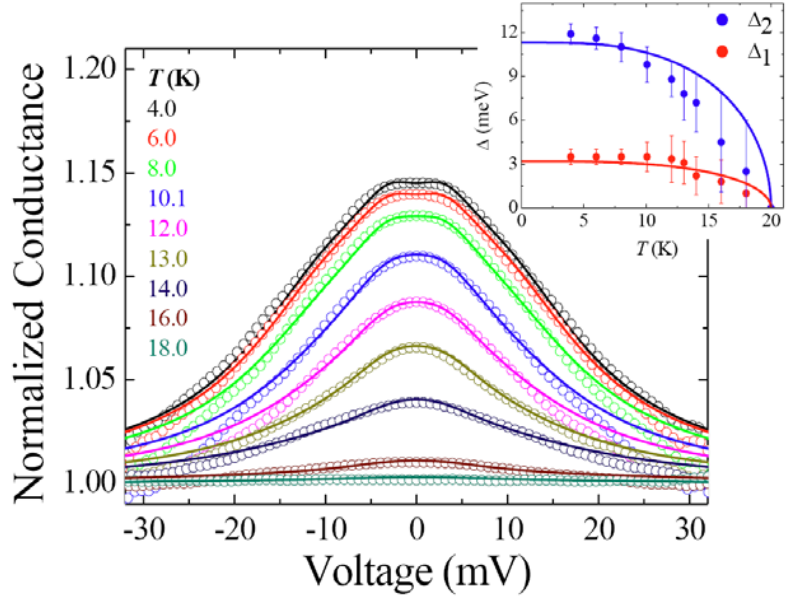

**FIG. 9.** (Color online) Temperature dependence of the normalized conductance curves (symbols) for the Ag/SmFeAs$_{0.5}$P$_{0.5}$O point contact with their relevant two-band BTK fitting curves (solid lines). The two-band fitting parameters are: $\Delta_1$ = 3.5 meV, $\Gamma_1$ = 3.4 meV, $Z_1$ = 0.29, $\Delta_2$ = 11.9 meV, $\Gamma_2$ = 9.15 meV, $Z_2$ = 0.225, $w_1$ = 0.3. Right inset: temperature behavior of the two gap values (symbols) as obtained by fitting the experimental curves. Two BCS trends are shown for comparison (solid lines).

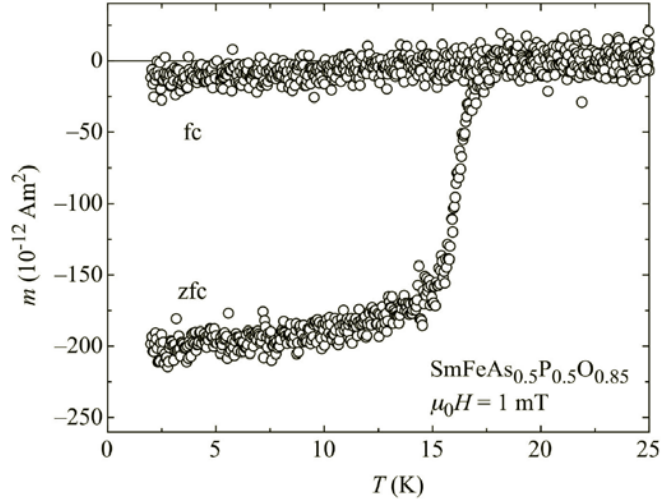

**FIG. 10.** The temperature dependence (zero-field-cooled and field-cooled) of the magnetic moment of a SmFeAs$_{0.5}$P$_{0.5}$O$_{0.85}$ (nominal composition) single crystal in a magnetic field of 1 mT applied parallel to the *c*-axis.

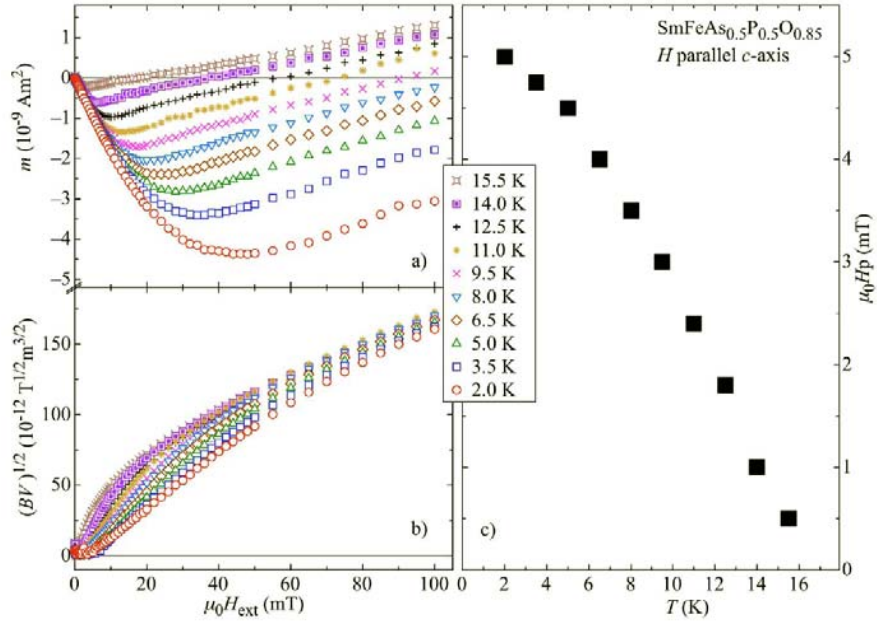

**FIG. 11.** (Color online) (a) Initial magnetization curves of a SmFeAs$_{0.5}$P$_{0.5}$O$_{0.85}$ crystal recorded in the temperature range between 2 and 17 K. (b) The quantity $(BV)^{1/2}$ calculated from the magnetic moment and plotted as a function of magnetic field. (c) The temperature evolution of the first penetration field $H_p$. The estimate $\mu_0 H_p(0) \approx 5$ mT yields an in-plane magnetic penetration depth $\lambda_{ab}(0) \approx 200$ nm.

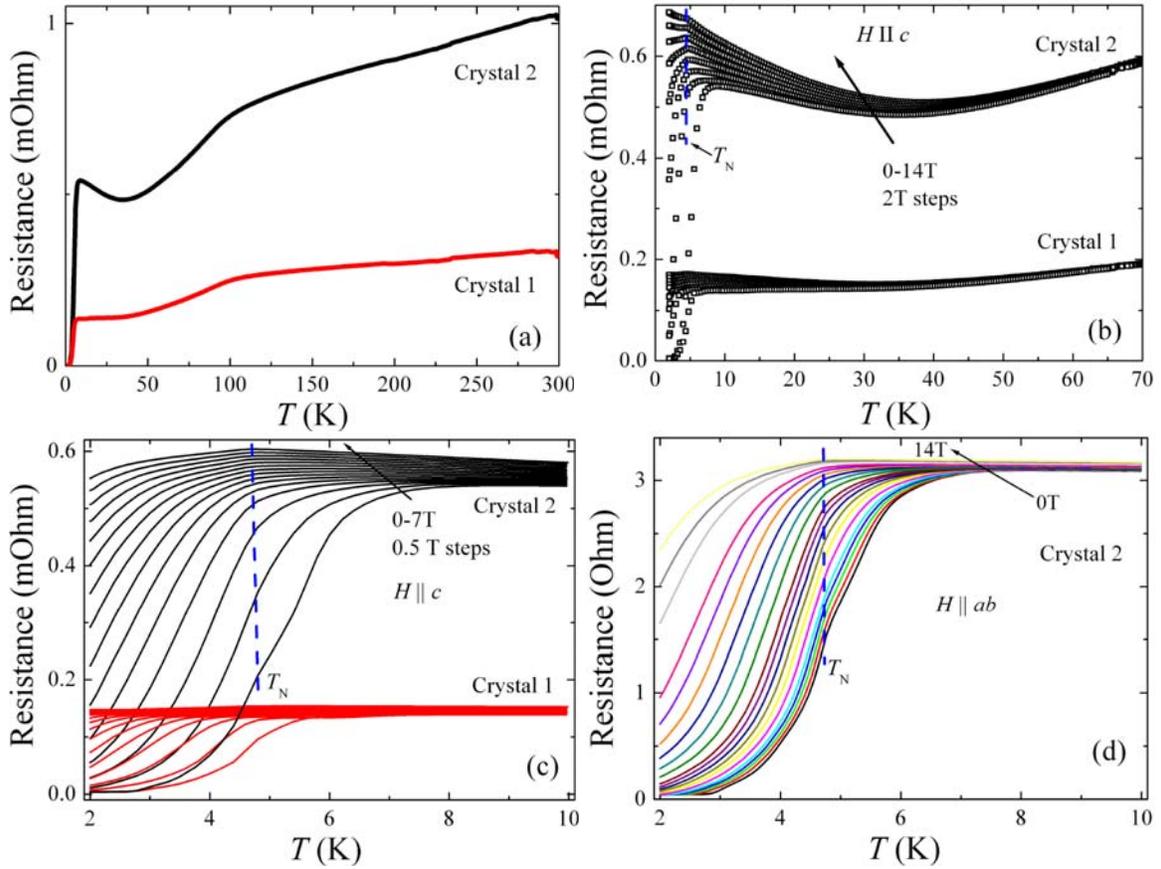

**FIG. 12.** (Color online) Temperature dependence of the resistance, in various fields, applied parallel to the $Fe_2(As,P)_2$ layers ($H \parallel ab$) and perpendicular to them ($H \parallel c$) for two $SmFeAs_{0.5}P_{0.5}O$ (nominal composition) crystals grown under high pressure. The dashed lines indicate the magnetic ordering temperature of Sm.

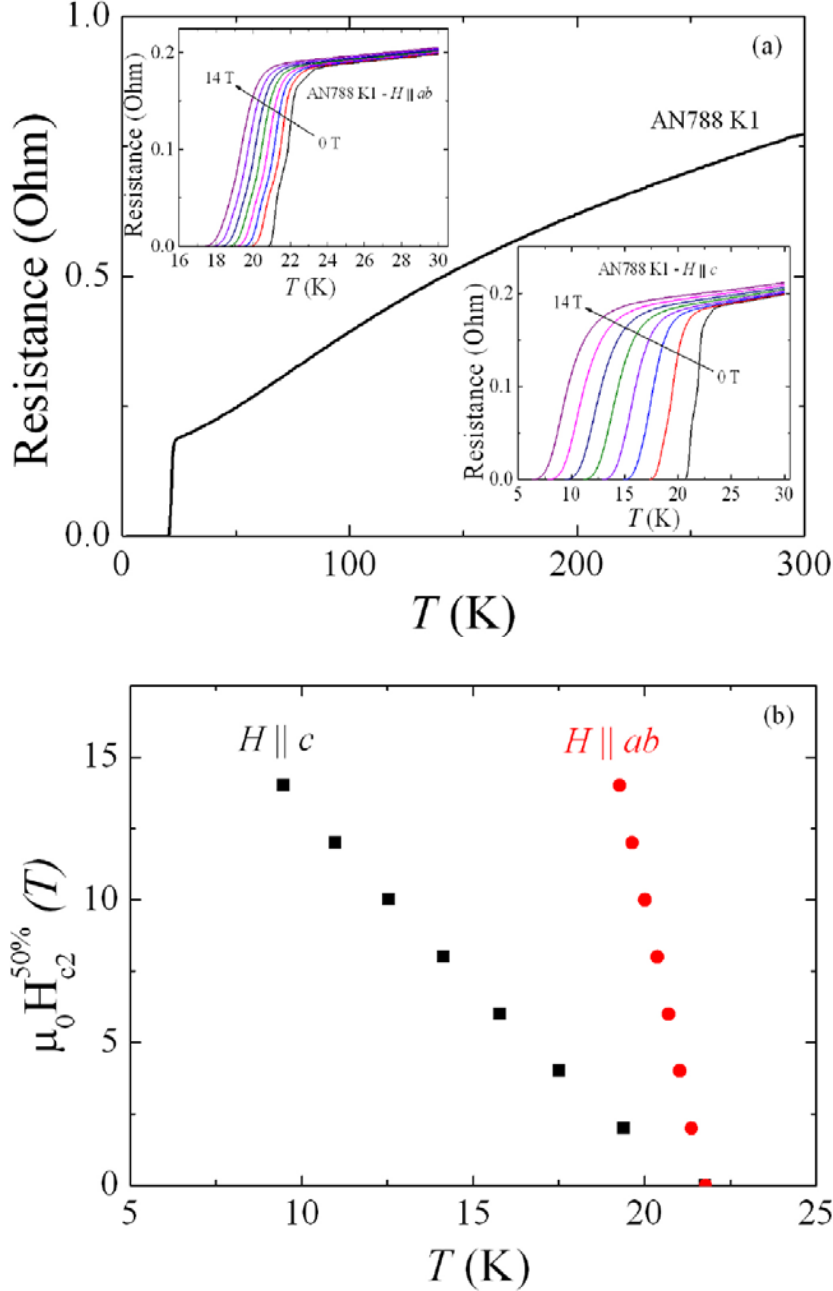

**FIG. 13.** (Color online) (a) Temperature dependence of the resistance for a SmFeAs$_{0.5}$P$_{0.5}$O$_{0.85}$ (nominal composition) single crystal. Left and right insets show measurements with the field applied parallel to the Fe$_2$(As,P)$_2$ layers ($H \parallel ab$) and perpendicular to them ($H \parallel c$), in various magnetic fields (0, 2, 4, 6, 8, 10, 12, 14). (b) Temperature dependence of the upper critical field with $H \parallel ab$ and $H \parallel c$. To determine $H_{c2}$ the 50% $\rho_n$ criterion was used.

**TABLE I.** Refined lattice constants, atomic parameters, and selected bond lengths (Å) and angles (deg) at room temperature for SmFeAs$_{1-x}$P$_x$O samples from the Rietveld refinements of the x-ray powder-diffraction profiles. The lattice is tetragonal with space group *P4/nmm*. Estimated errors in the last digits are given in parentheses.

| Nominal composition Abbreviation | SmFeAsO AP | SmFeAs$_{0.44}$P$_{0.56}$O AP | SmFeAs$_{0.5}$P$_{0.5}$O AP | SmFeAs$_{0.5}$P$_{0.5}$O AP+HP | SmFeAs$_{0.5}$P$_{0.5}$O HP | SmFeAs$_{0.5}$P$_{0.5}$O HP+AP | SmFeAs$_{0.5}$P$_{0.5}$O HP+AP+HP | SmFePO AP |
|---|---|---|---|---|---|---|---|---|
| $T_{c,eff}$ (K) | NO SC | NO SC | NO SC | 10.2 | 23.4 | NO SC | 18.0 | NO SC |
| $a$ (Å) | 3.92520(4) | 3.90225(4) | 3.90422(4) | 3.90655(6) | 3.90664(2) | 3.90964(4) | 3.9059(1) | 3.87431(8) |
| $c$ (Å) | 8.4693(1) | 8.3111(1) | 8.3206(1) | 8.3181(2) | 8.3489(4) | 8.3369(1) | 8.3363(3) | 8.1912(2) |
| $V$ (Å$^3$) | 130.488(3) | 126.558(2) | 126.831(3) | 126.943(4) | 127.419(9) | 127.431(3) | 127.179(7) | 122.952(5) |
| $z_{Sm}$ | 0.1371(2) | 0.1417(3) | 0.1412(3) | 0.1418(2) | 0.1416(4) | 0.1411(2) | 0.1416(4) | 0.1466(4) |
| $B_{iso}$ [a] (Å$^2$) | 1.37(9) | 2.25(7) | 1.58(8) | 3.84(8) | 2.4(2) | 2.23(7) | 2.4(2) | 1.80(9) |
| Fe $B_{iso}$ (Å$^2$) | 2.3(2) | 2.62(2) | 2.35(18) | 5.3(2) | 4.1(4) | 2.7(2) | 4.3(3) | 2.02(2) |
| $z_{As/P}$ | 0.6618(4) | 0.6537(7) | 0.6537(7) | 0.6586(7) | 0.658(1) | 0.6536(7) | 0.656(1) | 0.643(2) |
| $B_{iso}$ (Å$^2$) | 1.2(2) | 2.2(2) | 1.6(2) | 4.2(2) | 3.0(5) | 2.3(2) | 3.9(4) | 2.3(3) |
| Occupation. As/P | 1.00/0.00(2) | 0.44/0.56(1) | 0.50/0.50(2) | 0.48/0.52(1) | 0.60/0.40(3) | 0.52/0.48(1) | 0.62/0.38(2) | 0.00/1.00(1) |
| O $B_{iso}$ (Å$^2$) | 1.2(7) | 3.5(6) | 1.2(5) | 4.6(6) | 1.6(9) | 2.9(6) | 3.0(9) | 3.7(9) |
| $R_p$ (%) | 1.65 | 1.85 | 1.68 | 1.11 | 1.53 | 1.37 | 1.54 | 1.77 |
| $R_{wp}$ (%) | 2.17 | 2.35 | 2.17 | 1.51 | 1.92 | 1.73 | 1.94 | 2.33 |
| $R_B/R_F$ | 3.94/4.81 | 5.28/6.21 | 5.61/6.39 | 7.64/6.61 | 5.11/4.48 | 6.40/5.68 | 5.42/4.94 | 13.4/15.4 |
| Sm-As/P (Å) | 3.256(2) | 3.241(3) | 3.246(3) | 3.223(3) | 3.230(5) | 3.251(3) | 3.236(5) | 3.238(8) |
| Sm-O (Å) | 2.2804(9) | 2.279(1) | 2.278(1) | 2.2818(9) | 2.283(2) | 2.2815(9) | 2.282(2) | 2.279(2) |
| Sm$_1$-Sm$_2$ (Å) | 3.619(2) | 3.628(2) | 3.625(2) | 3.633(2) | 3.636(3) | 3.630(2) | 3.633(3) | 3.643(3) |
| Sm$_2$-Sm$_3$ (*a*) (Å) | 3.92520(4) | 3.90225(4) | 3.90422(4) | 3.90655(6) | 3.90664(2) | 3.90964(4) | 3.9059(1) | 3.87431(8) |
| Fe-As/P (Å) | 2.394(2) | 2.332(3) | 2.334(3) | 2.357(3) | 2.357(6) | 2.337(3) | 2.347(5) | 2.262(8) |
| O-O=Fe-Fe (Å) | 2.7755 | 2.7593 | 2.7607 | 2.7624(4) | 2.7624 | 2.7645 | 2.7619 | 2.7396 |
| As$_1$-As$_2$ (Å) | 3.901(3) | 3.760(6) | 3.764(6) | 3.820(6) | 3.82(1) | 3.769(6) | 3.795(9) | 3.601(14) |
| As$_2$-As$_3$ (*a*) (Å) | 3.92520(4) | 3.90225(4) | 3.90422(4) | 3.90655(6) | 3.90664(2) | 3.90964(4) | 3.9059(1) | 3.87431(8) |
| As$_1$-Fe-As$_2$, $\beta$ (deg) | 109.13(15) | 107.5(2) | 107.5(2) | 108.3(3) | 108.2(4) | 107.5(2) | 107.9(4) | 105.5(6) |
| As$_2$-Fe-As$_3$, $\alpha$ (deg) | 110.15(7) | 113.6(1) | 113.54(11) | 111.93(1) | 111.97(19) | 113.54(11) | 112.65(18) | 117.8(3) |
| S$_3$ [b] (Å) | 1.703(4) | 1.700(6) | 1.707(6) | 1.660(6) | 1.673(8) | 1.712(6) | 1.687(9) | 1.72(2) |
| S$_1$ [b] (Å) | 2.322(3) | 2.355(5) | 2.350(5) | 2.359(3) | 2.364(7) | 2.353(3) | 2.361(7) | 2.402(7) |
| $h_{Pn}$ (S$_2$/2) [b] (Å) | 1.370(3) | 1.277(6) | 1.279(6) | 1.319(6) | 1.319(8) | 1.281(6) | 1.300(8) | 1.17(2) |

[a] Debye-Waller factor. [b] Figure 3(a).

**TABLE II.** Crystallographic and structural refinement parameters for a SmFeAs$_{0.5}$P$_{0.5}$O$_{0.85}$ single crystal. (T = 295 K, Mo Kα, λ = 0.71073 Å). The absorption correction was done analytically. A full-matrix least-squares method was employed to optimize $F^2$. Some distances and marking of atoms are shown in Fig. 3 and 4.

| Crystallographic formula | SmFeAs$_{0.53}$P$_{0.47}$O$_{1-y}$ |
|---|---|
| $T_{c,eff}$ (K) | 17 |
| Unit cell dimensions (Å) | $a$= 3.91022(8), $c$= 8.3367(2) |
| Volume (Å$^3$) | 127.467(5) |
| $z_{Sm}$ (atomic coordinate) | 0.1414(1) |
| $z_{As/P}$ | 0.6551(2) |
| Sm$_1$-Sm$_2$ (Å) | 3.6342(4) |
| O-O = Fe-Fe (Å) | 2.7649(2) |
| Sm$_2$-As$_2$ (Å) | 3.2438(7) |
| Sm-O (Å) | 2.2831(2) |
| As$_1$-As$_2$ (Å) | 3.7863(14) |
| Fe-As/P (Å) | 2.3440(7) |
| As$_1$-Fe-As$_2$, β (deg) | 107.72(3) |
| As$_2$-Fe-As$_3$, α (deg) | 113.03(3) |
| $S_3$ (Å) | 1.697(2) |
| $S_1$ (Å)     Fig.3(a) | 2.358(2) |
| $h_{pn}$ ($S_2$/2) (Å) | 1.293(2) |
| Calculated density (g/cm$^3$) | 7.297 |
| Absorption coefficient (mm$^{-1}$) | 35.630 |
| $F$(000) | 244 |
| Crystal size, (μm$^3$) | 117 × 77 × 18 |
| θ range for data collection | 2.44° - 48.39° |
| Index ranges | -7≤$h$≤7, -8≤$k$≤7, -16≤$l$≤17 |
| Reflections collected/unique | 3673/409 $R_{int.}$ = 0.0536 |
| Completeness to 2θ | 98.3 % |
| Data/restraints/parameters | 409/0/13 |
| Goodness of fit on $F^2$ | 1.135 |
| Final $R$ indices [$I$>2σ($I$)] | $R_1$ = 0.0324, w$R_2$ = 0.0742 |
| $R$ indices (all data) | $R_1$ = 0.0410, w$R_2$ = 0.0767 |
| Δρ$_{max}$ and Δρ$_{min}$, (e/A$^3$) | 4.238 and -4.408 |